\documentstyle[aps,preprint]{revtex}
\input psfig.tex

\def\nhfcls{{\rm NH}_4{\rm Cl}_{\rm (s)}}

\def\nhfp{{\rm NH}_4^+}

\def\ndfcls{{\rm ND}_4{\rm Cl}_{\rm (s)}}
\def\chloride{{\rm Cl}^-}

\begin{document}

\title{Monte Carlo Studies
of the Orientational Order-Disorder Phase Transition
in Solid Ammonium Chloride }

\author{Robert Q. Topper$^*$ and David L. Freeman}
\address{Department of Chemistry, University of Rhode Island, Kingston,
RI 02881}

\renewcommand\baselinestretch{1}
\addtocounter{footnote}{1}
\footnotetext{Present Address: Department of Chemistry, The Cooper
Union for the Advancement of Science and Art,
Cooper Square, New York, NY 10003}
\renewcommand\baselinestretch{1.5}

\maketitle


\begin{abstract}
Monte Carlo methods are used to study
the phase transition in $\nhfcls$
from the orientationally ordered $\delta$ phase to the orientationally
disordered
$\gamma$ phase.  An effective pair potential is used to model
the interaction between ions.
Thermodynamic properties are computed in the canonical and
isothermal-isobaric
ensembles.  Each ammonium ion is treated as a rigidly rotating body
and the lattice is fixed in the low-temperature CsCl geometry.
A simple extension of the Metropolis Monte Carlo method
is used to overcome quasiergodicity in the rotational sampling. In the
constant-$NVT$ calculations the lattice is held rigid; in the constant-$NpT$
calculations
the lattice parameter is allowed to fluctuate.  In both ensembles
the order parameter rapidly falls to zero in the range
(200 - 250)K, suggesting that the model disorders at a temperature in
fair agreement with the experimental disordering temperature (243K).
Peaks in the heat capacity and thermal expansivity curves are also found
in the same temperature range.
\end{abstract}
\pacs{PACS numbers: }


\section{INTRODUCTION AND EXPERIMENTAL SUMMARY}

The general study of phase transitions has been of continuing
 interest to condensed matter chemistry
and physics for many years.  Particular efforts have been made to
characterize and understand second-order (or continuous)
phase transitions and associated critical phenomena~\cite{goldenfeld}.
  Good
examples of such continuous phase transitions are found in certain
atomic and molecular solids when they undergo order to disorder
transitions~\cite{disorder}.  These order-disorder transitions are
probed experimentally by observing anomalies in thermodynamic properties
like heat capacities and various susceptibilities.
  Some examples of systems that are known
to exhibit disordering phenomena are $\nhfcls$, $\beta$-brass, and
 ${\rm KCN}_{({\rm s})}$.

In this study we focus on the order-disorder phase transition
in $\nhfcls$, an interesting case that has received considerable
previous
theoretical and experimental
attention~\cite{disorder}.  A helpful
review of the experimental data up to 1978
is given
by Parsonage and Staveley~\cite{disorder}.
As discussed in Reference 2 ammonium chloride has three
phases.
The two phases generally referred to as $\delta$ and $\gamma$
share the CsCl lattice type
shown in Figure 1, whereas the $\alpha$ phase has a NaCl lattice
type.  First measured and reported by Simon in 1922~\cite{Simon}
and subsequently by other workers\cite{ber,TrappVander,VorGar,Schwartz}, the
$\delta \rightarrow \gamma$ transition has been used as a textbook example
of an order-disorder phase transition~\cite{experpchem}.
{}From experiment it is known that in the low-temperature $\delta$ phase,
most of the $\nhfp$ molecules are in the same orientation, (
the ammonium ions orient with their hydrogen atoms pointing
towards the ``same" crystallographic axes),  and the $\delta$ phase
is said to be ``orientationally ordered.''  In
the $\gamma$ phase, the orientations with respect to the chloride ions
have no long range order~\cite{disorder}.  This model of the
$\delta \rightarrow \gamma$ phase transition
has been supported by various spectroscopic
measurements, including neutron scattering experiments and NMR
studies~\cite{WagHorn,LevyPet,ManTrapp,TopRichtSpring,BornHohlEck,Mukhetal}.

The origin of the orientational order in NH$_4$Cl can be understood in terms
of the intermolecular forces of the crystal.  There are attractive
interactions between the hydrogen atoms in an ammonium ion and the
adjacent cage of chloride ions (see Figure 1).  Any isolated ammonium
ion in a chloride cage tends to align with either of two
equivalent sets of four chloride ions.  In Figure 1 the hydrogen atoms
point toward the set of chloride ions that has been shaded.
  In
addition to the interactions of ammonium ions with the near-neighbor
chloride cage, the ammonium ions interact with each other
electrostatically, with the lowest-order, non-zero contribution to the
interaction arising
 from the
octapole-octapole interactions characteristic of tetrahedral charge
distributions.  The octapole-octapole interaction has an asymptotic decay
of $R^{-7}$ ($R$ is the distance between the nitrogen centers on any two
ammonium ions). There is an absolute potential energy minimum if two
interacting ammonium ions are aligned,
and there is a higher energy local minimum if two ammonium ions have
opposite alignments.
We can
think of the two potential minima for the ammonium ion in a chloride
cage as two available states.  These two available states can be mapped
onto a two-state Ising model by identifying one orientation with
an up spin and the other possible orientation with a down
spin\cite{expspin}.
The
order-disorder transition has often been interpreted~\cite{isingf}
 in terms of such a two
state Ising model with a positive ferromagnetic coupling constant between near
neighbor spins.  We shall investigate the energetics of this Ising
picture within a simple pair site potential model in Section
\ref{s:model}
of this paper.

Although the interpretation of the transition in terms of the two-state
 Ising model has been insightful, it is not a perfect description
of the transition.
At atmospheric pressure, a measurement of the heat capacity $C_p$ as a function
of
temperature shows a distinctly $\lambda$-like shape with a peak at
243K\cite{Schwartz}.
  While this $\lambda$ character has been used to support
 the interpretation
of this transition in terms of the Ising model, it is known that
there is a discontinuous change in the specific volume at
this temperature~\cite{Fredericks,NisPoy,WeinGarl}, as well as a latent heat of
transition~\cite{TrappVander,VorGar}.  This transition
at atmospheric pressure is, at least,
weakly first order, not withstanding the qualitative similarity the system
has to an Ising lattice.  The first-order character of the transition
has led to descriptions of the system in terms of a compressible Ising
model\cite{isingf,rgar,baker,baker1,slichter}.

It has also been established that there is hysteresis associated with
the $\delta \rightarrow \gamma$ phase transition in measurements of
the specific volume~\cite{Fredericks,NisPoy},
the spin-lattice relaxation time~\cite{ManTrapp,TopRichtSpring}, and neutron
scattering measurements of the [111] Bragg
intensity~\cite{TopRichtSpring}.  Pressure-dependent
studies of this system~\cite{TrappVander,ManTrapp,WeinGarl}
show that the nature of the transition is a function of pressure.
At high pressures ( greater than 1400 atmospheres),
the transition changes from first order to continuous.  This
transformation in behavior is characteristic of a
tricritical point~\cite{goldenfeld}.  The location of the tricritical
point on the phase diagram is known to shift to lower pressures when the
ammonium ions are deuterated\cite{NisPoy,WeinGarl},
implying the importance of
quantum effects.
To date no complete explanation of the microscopic
origin of the tricritical point in ammonium chloride and its
deuterated counterparts is available.

The purpose of the work presented here is to provide information leading
to a
microscopic description of the order-disorder phenomena in ammonium
chloride that is more complete than the traditional Ising picture.
Our motivation for the study comes from recent
investigations\cite{berry}
into the analogues of phase transition phenomena in small
clusters.  Many small clusters have thermodynamic properties as a
function of temperature that have been interpreted as analogues of bulk
phase transition behavior.  As examples of
first-order transition behavior, small clusters of rare gas
atoms exhibit heat capacity anomalies\cite{ffd}, and rapid increases in
diffusivities\cite{berry} that are characteristic of bulk melting.
Recently, the study of phase transition analogues in small clusters has
been extended to cases that are similar to second-order transitions.
Lopez and Freeman~\cite{lopez}  observed
heat capacity
anomalies in model
Pd-Ni alloy clusters,
that are reminiscent of the order-disorder transitions known to occur in
some bulk alloy materials like $\beta$-brass\cite{kittel}.  Studies of
analogues to second-order phase transitions in other kinds of materials
continue to be of interest.  Given this background, it is our
intention ultimately
to investigate
the possibility of orientational disordering transitions in ammonium
chloride clusters. To do justice to such a study
a detailed simulation of the analogous bulk
transition using the same model potential is important.  Providing bulk
information for future work on the cluster systems is an important
motivation of the work we report.

A review of some previous computational studies of
transitions in molecular and
ionic crystals has been given by Klein~\cite{kleinrev}.
There have also been studies related to the present
work by Smith~\cite{smith}
and by H\"{u}ller and Kane~\cite{huller}
 who focused on the orientational motion of the
ammonium ions and by
O'Shea~\cite{shea} who studied related octapolar solids.  More pertinent
to the present discussion is the work
of Klein, McDonald and Ozaki~\cite{KleinMacOz}
 who studied NH$_4$Br, KClO$_4$ and
Li$_2$SO$_4$ using molecular dynamics methods.  Klein {\em et
al.}~\cite{KleinMacOz}
introduced a suitable order parameter to monitor the disordering
transition in ammonium bromide.  We shall make use of the same order
parameter in the work we report.

The contents and organization of the remainder of this paper are as
follows.  In the next section the model potential is discussed along
with the basic structural features of the ammonium chloride lattice
implied by the model.  In Section III we discuss the computational
details including how we resolve quasiergodicity problems in the
simulations.  We present the computed thermodynamic properties in
Section IV including a discussion of the chosen order parameter.  Of the
properties given in Section IV, we show that the model predicts no peak
in the isothermal compressibility at the disordering transition for the
simulation sample sizes used in the current work.  In Section V we
identify the lack of peak in the compressibility with finite size
effects by presenting results of thermodynamic properties predicted in a
compressible Ising calculation as a function of sample size.  We
summarize our findings and discuss their significance in Section VI.

\section{Theoretical Model} \label{s:model}

\subsection{Model Potential}

In the present study we use classical Monte Carlo methods to compute
thermodynamic
properties for $\nhfcls$ as it undergoes an order to disorder transition.  The
application of classical mechanics to
pair models of ionic solids has been explored by Klein and coworkers
{}~\cite{kleinrev,KleinMacOz},
who have
used molecular dynamics methods to study dynamical processes in
several ionic crystals.  Their work gives us
good reason
to expect that useful information about the order to disorder transition
in ammonium chloride can be obtained using a simple pair model potential
dominated by Coulombic contributions.
We independently evaluate the success of this approach
by using the model potential to predict such bulk properties
as the lattice constant
and the barrier to rotation of a single ammonium ion
 between the two local potential
minima.  We shall confirm that the model potential yields values for
these observables that are in reasonable
agreement with experiment, and we can expect that the potential is
sufficiently accurate
for our purposes.
The approach we use to validate the effective pair potential has been
advocated elsewhere~\cite{AllenTild}.

The specification of the potential surface is somewhat simplified by the
dynamical
model we assume.  In all calculations reported here, we treat the $\nhfp$
molecular
 ions
as rigidly rotating bodies, with the center of mass of each $\nhfp$ fixed at
the
center of
each cubic cell.  This dynamical model makes it unnecessary to specify an
internal vibrational potential
for the $\nhfp$ molecule.  We need only set $\nhfp$ - $\nhfp$,
$\nhfp$ - $\chloride$, and $\chloride$ - $\chloride$ interactions to
specify a useful effective interaction potential.
Each of these interactions is
further
 decomposed
into a total of six pair atom-atom interaction terms.  Letting {\bf r}
 designate the
coordinates of all interacting particles in the system under consideration,
we assume that the total interaction potential $U({\bf r})$ is given by

\begin{equation}
U({\bf r}) = \sum_i \sum_{j > i}
u_{2} (r_{ij})
\end {equation}
where $r_{ij}$ is the distance between particle $i$ and particle $j$, and
$u_{2}$ is
a pair potential of the form

\begin {equation}
u_2 (r_{ij}) = A_{ij} \exp(- \alpha_{ij} r_{ij})
          + {  {D_{ij} }\over {r_{ij}^{12}} }
          + {{q_i q_j }\over {r_{ij}}   }
          -   {  {C_{ij} \over {r_{ij}^6} } } .\label{eq:pair}
\end{equation}
Most of the parameters used to define the potential are presented in
Table 1.  In addition as used elsewhere~\cite{KleinMacOz,jorg1} we set
$q_H=.35$
and $q_N=-.40$.
  We also set $q_{Cl}=-1.00$~\cite{jorg}.  The charges used on the
nitrogen and hydrogen atoms of the ammonium ion were confirmed by
independent {\em
ab initio} calculations~\cite{ferg}.
In Table 1
we have assumed transferability of most of the atom-atom interaction
parameters from other, similar
systems studied previously by Klein {\em et al.}~\cite{KleinMacOz}
 and Pettit and Rossky~\cite{pr}. For
the parameters not available from previous work, we use
standard combination rules, i.e.
relations of the form
\begin{equation}
A_{ij} = \sqrt{A_{ii}A_{jj}} ,
\end{equation}
\begin{equation}
C_{ij} = \sqrt{C_{ii}C_{jj}} ,
\end{equation}
and
\begin{equation}
\alpha_{ij} = { {\alpha_{ii}+\alpha_{jj}}\over{2} } .
\end{equation}
The specific sources for the parameters used in Eq.(\ref{eq:pair}) are
given as footnotes in Table 1.

As in any calculation,
to generate the results that follow both in this and subsequent sections, a
finite representation of the lattice was used.  To reduce the
importance of edge
effects, we also used standard minimum image periodic boundary
conditions~\cite{AllenTild}.
In the main (or central) sample cell we included either 8
ion pairs (48 atoms) corresponding to a 2x2x2 lattice, or 27 ion pairs
(162 atoms)
corresponding to a 3x3x3 lattice.  To evaluate efficiently the sums over
the periodic images we also used standard Ewald methods with vacuum
boundary conditions~\cite{AllenTild}.
We found the Ewald calculations to be converged by
setting the decay parameter of the error functions to 5.55 in units of
the inverse length of a side of the central sample cell~\cite{heyes}
and including
125 reciprocal lattice vectors in the reciprocal lattice sum.

\subsection{Properties of the lattice within the model potential}
\label{s:prop}

A primary indication of the validity of the parameters listed
in Table 1 is given by comparing the lattice constant and cohesive energy
at 0K predicted
by the model potential with experiment.  Using the parameters in Table
1, we have minimized the energy of ammonium chloride using the cesium
chloride phase.  We have assumed all ammonium ions in the lattice are
oriented in the same direction with respect to the crystallographic axes
in the manner shown in Figure 1.  The number of ion pairs in the
central simulation cell was taken to be 27, and the Ewald sums have been
evaluated as discussed above.  The resulting lattice constant
$a_0$ (the
distance between the nitrogen atom on an ammonium ion and an adjacent
chloride ion) is calculated to be 3.789  \AA \ with a cohesive energy of
-759.7 kJ mol$^{-1}$.  The agreement with the experimental lattice
constant (3.868 \ \AA)~\cite{lbj}
 and the experimental cohesive energy at 298K (-697 kj
mol$^{-1}$)~\cite{wilson}
is within acceptable limits.  The differences
between the calculated and experimental values are a result of finite size
and thermal effects as well as the details of the model potential.

Another indication of the accuracy of the model potential is given in
Figure 2 where the potential energy $U$ of the lattice with 27 ion pairs
in the
central lattice is plotted as a function of the rotation angle $\phi$ of the
central ammonium ion [see Figure 1].  At an angle of $\phi=0$
(where we set the zero of energy in this figure)
all the ammonium ions in
the lattice are oriented with respect to a set of crytallographically
equivalent chloride ions.  The central ammonium ion is then rotated as
in Figure 1 until an angle of $\phi=\pi /2$ where the central ammonium
ion is oriented with respect to the alternate set of chloride ions in
the lattice.  At this angle the ammonium ion finds a local potential
minimum higher in energy than the absolute minimum at $\phi=0$.  The
$\sim$18 kJ mol$^{-1}$ barrier to rotation evident in Figure 2 is in
good agreement with experimental estimates based on NMR
and other data~\cite{disorder}.  This
agreement is another indication that the potential model can be expected to
be at least
sufficiently accurate to account
qualitatively for the order to disorder phenomenology
of the system.

In addition to calculating the barrier to rotation for the ammonium
ions, we investigated the effective range of the ammonium-ammonium
interactions in the lattice.  As an initial geometry we considered a
lattice with all ammonium ions oriented in the same direction with
respect to a set of crytallographically equivalent set of chloride ions
except for the central ammonium ion in the simulation cell.
This central ammonium ion was oriented to the other set of chloride ions (i.e.
at an angle of $\phi =\pi /2$ in Figure 2).
We then compared the energy of the lattice with a single ammonium ion
out of orientation, with the energy obtained by
rotating another ammonium ion in the lattice at a distance $R$ from the
central ammonium ion so that the central ion and the additional ion were
oriented in the same direction.
It is important to recognize that the energy calculations were performed
without any nearest neighbor assumptions about the range of the interactions.
The change in potential energy $\Delta U$
 obtained from this
process is given in Figure 3 as a function of $R$.
At distances beyond the first ammonium ion cage, $\Delta U$ becomes
constant,
and
as is evident from Figure
3, the interaction between ammonium ions can be well represented by
a nearest neighbor model.

\subsection{Implications}

The simple pair model expressed in Eq.(\ref{eq:pair}) gives rise to a stable
CsCl lattice with the ammonium ions oriented in the same direction in
its lowest energy 0K structure.  The orientational ordering is a
consequence of the attractive octapole-octapole interaction between the
ammonium ions.  For ammonium chloride the octapole-octapole
coupling is sufficiently weak that interactions beyond the near neighbors
can be neglected.

These preliminary indications lend support to representing the
system
by a simple Ising model with a positive ferromagnetic coupling
constant. However,
as indicated in the Introduction the simple Ising picture is
not of sufficient complexity to explain the phenomenology of the order
to disorder transition.  The transition from first order to second order
behavior, characteristic of a tricritical point, provides justification
for investigating the transition with more detail than the Ising
picture.  In the next section we develop the necessary Monte Carlo tools
to investigate the system within the model potential of
Eq.(\ref{eq:pair}).  By performing simulations in the
isothermal-isobaric ensemble, coupling between lattice motions and
rotational motions of the ammonium ions will be included.

\section{Simulation method}

The details of the Metropolis Monte Carlo method~\cite{mrrtt}
both in the canonical and in the isothermal-isobaric ensemble
been discussed in many references~\cite{AllenTild,KalWhit,Mac}.
In this section we explain some of the
details specific to the simulations performed in the current work, and
the approach we used to insure that the simulations were done in
an ergodic fashion.

Since the Ewald sums
constituted the dominant fraction of the computational effort in this
work, we examined the consequences of decreasing the number of reciprocal
lattice
vectors.  Although the absolute value of the cohesive energy was
sensitive to including fewer reciprocal lattice vectors,
we found the lattice constant
and the barrier to rotation of the ammonium ions (See Section
\ref{s:prop})
to change by $\sim$0.3 percent when only 8 reciprocal lattice vectors
(rather than the converged 125) were included.
  Our interest is in the order to disorder
transition, and the transition can be expected to be sensitive to the
barriers and not to the absolute value of the cohesive energy of the
crystal.  We tested this assumption by comparing fluctuation
quantities
(e.g. heat capacities, compressibilities, etc.)
sensitive to the location of the transition in the canonical
ensemble with 8 and 125 reciprocal lattice vectors included.  We observed no
significant changes in the computed properties.  Most of the Monte Carlo
results
were determined with the smaller set of
reciprocal lattice vectors.

The calculations in the canonical ensemble were performed using central
cell sizes of both 8 ion pairs and 27 ion pairs.  For each central cell
size, the lattice parameters were adjusted to give the minimum energy
for the orientation having
 all the ammonium ions in the same direction.
About each ammonium ion in the lattice we constructed a set of
orthogonal Cartesian axes.
Each
Monte Carlo point in the canonical simulations consisted of a rotation
of a randomly chosen ammonium ion about
one Cartesian axis randomly chosen~\cite{watt}.
  The
maximum allowed rotation angle was adjusted so that about fifty percent
of the moves were accepted.  As is typical in Monte Carlo simulations,
this maximum allowed rotation angle was a function of temperature and
was found by performing short run experiments at each temperature.  In
addition to these normal Metropolis moves, we found it necessary to
include moves with larger maximum displacements ten percent of the
time.  The purpose of the moves with magnified displacements (magwalking)
was to insure that the ammonium ions were given the opportunity to
overcome the potential barrier separating the two minima in the
potential surface (See Figure 2).  Consequently, the maximum
displacement in the magwalking moves was taken to be $\pi /2$.

Evidence that the magwalking scheme discussed in the previous paragraph
provides an ergodic distribution is given in Figure 4 where the average
potential
energy of the 8 ion pair lattice is plotted as a function of
temperature.  The data used in generating both curves of Figure 4 was
initiated with configurations having
 the ammonium ions in random orientations.  Quasiergodicity
difficulties~\cite{ffd,vw}
can be anticipated at low temperatures where all the
ammonium ions are expected to have the same orientation with respect to
the chloride ions.  The data in the upper curve of Figure 4 was
generated using Metropolis Monte Carlo methods with a fixed step size
for all moves.  The unstable behavior at low temperatures is evident.
The instability is a result of rapid quenching of the initial random
orientations of
the ammonium ions into
local, high-energy minima.  This disordered trapping causes the
quasiergodicity in the sampling of the rotations.
In the lower curve of Figure 4, the magwalking scheme described in
the previous paragraph was used.  The sensible behavior at low
temperatures is clear, and the simple magwalking scheme can be expected
to solve the quasiergodicity problems in this system.

In the isothermal-isobaric simulations the Monte Carlo sampling is with
respect to the distribution
\begin{equation}
\rho({\bf r})= \Delta ^{-1} \exp(-\beta U ({\bf r}) -\beta pV )
\end{equation}
where $U$ is the system potential energy, $\beta = 1/k_B T$ where $T$ is
the temperature and $k_B$ is the Boltzmann constant, $p$ is the
pressure,
 $V$ is the volume
and $\Delta^{-1}$ is a normalization.
  To perform the isothermal-isobaric simulations
 in addition to the rotational
moves used in the canonical study, the lattice parameter (and consequently
the system volume) was varied.
{}From numerical experiments we found that including volume fluctuations
about forty percent of the time gave good convergence of the computed
properties.
The volume fluctuations were included with Metropolis Monte Carlo moves
again with a maximum displacement chosen so that about fifty percent of
the attempted fluctuations were accepted.  Unlike the rotational moves,
we found no evidence of quasiergodicity in the volume fluctuations.
When a mixture of maximum step sizes were used in the volume
moves, we found no
statistically significant changes in the computed thermodynamic
properties.

The volume fluctuations included in this work correspond to
a single vibrational breathing mode for the lattice.  Of course, the
real lattice dynamics in NH$_4$Cl is more complex than this simple
picture.  However, at least we have been able to include important
contributions to the coupling between the rotational and vibrational
modes of the lattice.

In the calculations that follow, we have calculated several fluctuation
quantities in addition to the energy $E$ and enthalpy $H=E+pV$
 of the crystal.
These fluctuation quantities include the constant volume and constant
pressure heat capacities
\begin{equation}
C_V/k_B = \frac {3 N }{2} + \beta ^2 [ <U^2>-<U>^2] \label{eq:cv}
\end{equation}
and
\begin{equation}
C_p/k_B = \beta ^2[<H^2>-<H>^2] \label{eq:cp}
\end{equation}
the isobaric coefficient of thermal expansion
\begin{equation}
\alpha = \frac {1}{V} \left ( \frac {\partial V}{\partial T} \right
)_{N,p}
\end{equation}
\begin{equation}
 = (k_B T^2 V)^{-1} [<VH>-<V><H>] \label{eq:alpha}
\end{equation}
and the isothermal compressibility
\begin{equation}
\kappa=-\frac {1}{V} \left ( \frac {\partial V}{\partial p} \right
)_{N,T}
\end{equation}
\begin{equation}
 = (k_B T V)^{-1} [<V^2>-<V>^2] \label{eq:kappa}
\end{equation}
In Eq.(\ref{eq:cv}) the notation $<>$ represents averages in the
canonical ensemble, and in Eqs.(\ref{eq:cp}),(\ref{eq:alpha})
and (\ref{eq:kappa}) the notation
$<>$ represents averages in the isothermal-isobaric ensemble.  In the
results we report in the next section, $C_V, C_p, \alpha$ and $\kappa$
were calculated directly from these fluctuation expressions.

In the calculations that follow, error bars were estimated at the double
standard deviation level by ``binning'' the data and estimating the
variance of the bin averages about the total walker average.  In the
limit of large numbers, this method is known to be appropriate for the
correlated data generated by a Metropolis walk.  However, for finite
Monte Carlo samples the suitability of the binning parameters should be
checked for a given model system to ensure the reliability of the error
estimates.  We performed these checks by comparing the computed error
estimates to those obtained through covariance calculations of the
error~\cite{stam,top}.
We observed the two methods to give similar results, indicating
that suitable binning parameters were chosen for this study.

\section{Results}

In this section we provide results of the Monte Carlo studies of the
properties of the NH$_4$Cl lattice as a function of temperature.  As
expected we shall find features in the thermodynamic properties as a
function of temperature that can be associated with a transition from
rotational order to disorder.  To monitor the degree of
 orientational order in the
lattice, we use an order parameter introduced by Klein {\em et al}~\cite
{KleinMacOz}.
To
define the order parameter we place the origin of a set of Cartesian axes
on the nitrogen atom of each ammonium ion, and we orient the Cartesian
axes so that the $z$-axis is orthogonal to
a square face of chloride
ions in the chloride cage.  The $x$ and $y$ axes are then oriented along
the edges of the lattice as shown by the coordinates displayed in Figure
1.
  For each ammonium ion we define
\begin{equation}
M_j=\frac {3 \sqrt{2}}{4} \sum_{i=1}^4 x_j^i y_j^i z_j^i \label{eq:mj}
\end{equation}
where $x_j^i$ is the $x$-component of the coordinate of
a unit vector pointing from the origin toward
the $i$'th
hydrogen atom on ammonium ion $j$ and the summation runs over the 4
hydrogen atoms on the ammonium ion.  $M_j$ is defined so that
$M_j=1$
when
ammonium ion $j$ is oriented exactly to one set of chloride ions, and
$M_j=-1$ when it is oriented exactly to the alternate set of chloride
ions (see Figure 1).
  Of course $M_j$ is a continuous function of the coordinates, so
that it takes on values of $\pm$ 1 only when the hydrogen atoms point
exactly to a set of chloride ions.  This definition of $M_j$ enables a
mapping of the orientations of the ammonium ions onto a spin variable.
When $M_j$ is positive we can think of ammonium ion $j$ as having a
positive or ``up'' spin and when $M_j$ is negative, we can think of
ammonium ion $j$ as having a ``down'' or negative spin.  To clarify the
mapping,
in Figure 5 we display a particular configuration of the ammonium chloride
lattice
taken from a canonical simulation with 8 ion pairs in the central
cell at 300K.  We have placed arrows on each nitrogen atom in the
lattice to carry information about the algebraic sign of $M_j$.
Positive $M_j$ is represented by an up arrow and can be interpreted as
an up spin.  Similarly, negative $M_j$ is represented by a down arrow and
can be interpreted as a down spin.
To simplify the
discussion often we shall describe the orientations of the ammonium ions in
terms of these spin variables.  In
analogy with the Ising model, we can also define the {\em magnetization per
site}
$M$ of the lattice by
\begin{equation}
M=\frac {1}{N} \sum_{j=1}^N M_j
\end{equation}
where the summation on $j$ is over the $N$ ammonium ions in the lattice.
Associated with the magnetization is a susceptibility per site $\chi$
defined by
\begin{equation}
\chi = N(<M^2>-<M>^2) \label{eq:chi}
\end{equation}
where the notation $<>$ in Eq.(\ref{eq:chi}) represents an ensemble
average.  We use $M$ as the order parameter for the orientational order to
disorder
transition in the system.

\subsection{Canonical simulations}

In the canonical simulations for each temperature we included 50000
passes
without the accumulation of data followed by 200000 passes with data
accumulation.  Each pass consisted of cycling through the ammonium ions
in the central simulation cell and attempting to rotate each ion once.

We begin by examining the changes in computed thermodynamic properties
that accompany alterations in the size of the central sample cell.  In
Figure \ref{f:cv} we present the constant volume heat capacity $C_V$
per ion pair in
units of the Boltzmann constant as a function of temperature.  The upper
curve gives results when the central cell consists of 8 ion pairs and the
lower curve gives results for a 27 ion pair central cell.  By increasing
the size of the central cell the width of the heat capacity maximum
narrows as expected.

By examining configuration files it is possible to verify that the
maxima in the heat capacity seen in Figure \ref{f:cv} are a result of
the rotational disordering of the ammonium chloride lattice.  We can
confirm this interpretation by monitoring the magnetization as a
function of temperature.  In Figure \ref{f:m2} the curve connected by
diamonds are computed values of $<M^2>$ as a function of temperature for
the 27 ion pair lattice.  We have chosen to present $<M^2>$ rather than
$<M>$ because in any finite lattice $<M>=0$ at any finite
temperature.  In a completely
oriented configuration as the temperature approaches zero, $M^2$ is
always 1 whereas $M$ can be either 1 or -1.  Then at low temperatures
$<M^2>$ must approach unity, and at high temperatures $<M^2>$ must
approach zero.  This behavior is evident in Figure \ref{f:m2}.  At the
transition temperature of $\sim$210K $<M^2>$ changes rapidly.  This
transition temperature matches the peak of the maximum in $C_V$ as a
function of temperature.  Also presented in Figure \ref{f:m2} is
the result of mapping the
magnetization onto a spin model.  The data for this spin model are
plotted in Figure \ref{f:m2} as points represented by triangles.  In the
spin model we set
\begin{equation}
M_s=\frac{1}{N} \sum_{j=1}^{N} \mbox{sign} (M_j)
\end{equation}
where sign($M_j$) is the algebraic sign of $M_j$ and $M_s$ represents
the magnetization of the system described by the spin variables.

  The difference
between the two curves given in Figure \ref{f:m2} clarifies the extent
to which the decrease in the magnetization can be attributed to
librational modes.  Since $M$ is a continuous function of the
coordinates of each ammonium ion, the magnetization will decrease as a
function of temperature even if all the ions remain ordered.  In
contrast, $M_s$ will change only if an ammonium
ion in a particular configuration
changes its ``spin;''
i.e. overcomes the barrier between the two minima in the rotational
potential surface. By comparing the decay of the order parameter to the decay
of the
Ising-like spin parameter in Figure \ref{f:m2}, we observe
that the initial decay of the order parameter is a result of the onset of
librational motions. At the transition temperature, the order parameter
rapidly decays to zero because of the loss of long range order of the
NH$_4^+$ orientations.

Although in a finite system $<M_s>$=0,
 in an actual Monte Carlo simulation $<M_s>$ may differ from
zero, because both inverted configurations may be reachable only in
Metropolis walks that are sufficiently long.  The length of the walk required
to actually calculate a zero value for $<M_s>$ will grow both with
decreasing temperature and increasing system size.  The effect of the
finite walks can be made apparent by examining the susceptibility $\chi$
as a function of temperature.  Such a graph is given in Figure
\ref{f:chi} for the lattice consisting of 27 ion pairs in the central
cell.  The maximum of $\chi$ occurs at the same temperature as the
temperature at which $<M^2>$ changes rapidly.  The fluctuations in
$\chi$ at this temperature are also large.  Of course in the limit of an
infinite system $<M> \neq 0$, and the susceptibility we calculate should
approach the infinite system result with increasing system size.

\subsection{Isothermal-isobaric simulations}

Canonical simulations do not have sufficient flexibility to incorporate
coupling of the rotational modes of the ammonium ions with the
vibrational modes of the lattice.  To obtain preliminary understanding
of the effects of such couplings, we have performed Monte Carlo
simulations of the thermodynamic properties of the system in the
isothermal-isobaric ensemble.  By fixing the pressure of the system,
the isothermal-isobaric simulations more closely match the experimental
situation than the canonical studies.  In exchange for this closer
connection with experimental data, calculations in the
isothermal-isobaric ensemble are computationally more demanding than
calculations in the canonical ensemble.  The range of system sizes that
can be studied while achieving acceptable levels of convergence is
restricted.

The simulations in the isothermal-isobaric ensemble were performed on a
27 ion pair representation of the ammonium chloride lattice in the
central cell.  At each temperature 300000 Monte Carlo passes were
performed without data accumulation followed by about one million passes
with data accumulation.  Each pass consisted of sequential attempted
rotations of each ammonium ion in the same manner as in the canonical
simulations.  Additionally, volume fluctuations were attempted in forty
percent of the passes.  Ten percent of the rotational moves used a
maximum rotation angle of $\pi /2$ using the magwalking scheme found to
be successful in the canonical simulations.  The remaining ninety per
cent of rotational moves used a maximum rotational displacement
determined so that about fifty percent of the attempted moves were
accepted.

In Figure \ref{f:m2npt} we present
$<M^2>$ as a function of temperature calculated both using the
definition of the order parameter given in Eq.(\ref{eq:mj}) (the points
represented by diamonds) and the projection of the order parameter onto
spin variables (the points represented by triangles).
The projected spin order parameter is the same as that used in the
canonical simulations and discussed previously.  The region of rapid
change in the order parameter is at a temperature of about 250K.  The
onset of the transition is more clearly seen in Figure \ref{f:chinpt}
where we plot the susceptibility as a function of temperature.  The
maximum occurs at a temperature of about 250K, a temperature where the
fluctuations in $\chi$ are large.

We measure the average volume of the crystal in terms of the lattice
parameter $a_0$.  The average volume is
of interest because we would expect to observe a rapid volume change
near any first order transition.  In Figure \ref{f:a0npt} the lattice
parameter in \AA \ is plotted as a function of temperature.  Although no
rapid change in the lattice parameter is seen, there is a clear
change of slope at the transition temperature.

In Figure \ref{f:cpnpt}  we give the constant pressure heat capacity
$C_p$ per ion pair,
the isothermal compressibility $\kappa$ and the isobaric
coefficient of thermal expansion $\alpha$ as a function of temperature.
The onset of the maxima in $C_p$ and $\alpha$ match the onset of the
order to disorder transition as identified by the order parameter.  From
basic considerations~\cite{whee}
a maximum is expected in $\kappa$ at the transition
temperature as well.  As is evident from Figure \ref{f:cpnpt}, $\kappa$
is found to be a monatomic increasing function of the temperature with
no apparent peak.
To test the sensitivity of the behavior of $\kappa$ to the number of
reciprocal lattice vectors included in the Ewald sums, we increased the
set from 8 to 125 reciprocal lattice vectors.  No significant change in the
compressibility as a function of temperature was observed.
We believe that the lack of peak is a consequence of
the finite size of the central simulation cell, and we give evidence for
this in the next section.

\section{A Compressible Ising Model}

In the previous section we showed that in the
isothermal-isobaric ensemble at the disordering transition,
peaks were observed in the constant
pressure heat capacity and in the isobaric coefficient of thermal
expansion.  The location of the peak maxima were in good agreement with
the temperature at which there were rapid changes in the order
parameter.  The identification of the peak maxima with the disordering
phenomena seems well justified.

In contrast to the properties discussed in the previous paragraph, no
maximum was observed in the isothermal compressibility.
Since the compressibility should have a specific heat-like divergence at the
transition temperature in the infinite system~\cite{whee},
the lack of observation is a concern.  In this section we
show that similar behavior is found in finite representations of a
compressible two-dimensional Ising model
on a square lattice, and the lack of peak in the
isothermal
compressibility we observed for NH$_4$Cl may be attributable to finite size
effects.

The compressible model we use is based on the two-dimensional Hamiltonian
\begin{equation}
H=-\frac{a}{R^7} \sum_{<ij>} S_i S_j + 4 \epsilon \left [ \left ( \frac
{\sigma}{R} \right )^{12}
- \left ( \frac {\sigma}{R} \right ) ^6 \right ] \label{eq:iham}
\end{equation}
where $S_i$ is the spin on site $i$ and can take on values of + or - 1,
the notation $<ij>$ on the sum represents summation over nearest
neighbors only, $a$, $\epsilon$ and $\sigma$ are parameters and $R$ is
the lattice constant.
The first term in the Hamiltonian is the standard spin-spin interaction
in the Ising model with a lattice-parameter dependent coupling constant.
We choose the $R$-dependence of the coupling constant to decay with
distance like the octapole-octapole interaction in ammonium chloride
($R^{-7}$).  The second term in the Hamiltonian is in the form of a
usual Lennard-Jones potential and provides a balance for the volume
fluctuations so that the lattice will relax to a physical nearest
neighbor distance.
In the Ising simulations that follow we have
attempted to choose parameters
for Eq. (\ref{eq:iham})
that mimic the ammonium chloride results
of the previous section.
For the bulk Lennard-Jones $\sigma$ parameter, we take a value that can
make the equilibrium lattice distance near that of the real crystal.  We
then take $\sigma=R_e/2^{1/6}$ where $R_e=7.31$ Bohr. We take $\epsilon$
to match the ammonium chloride lattice energy of 8380K.
Since the barrier to changing the orientation
of a single ammonium ion in a completely ordered ammonium chloride
lattice is about 1000K, we take $a=2000 R_e^{7}/N_{\mbox{spin}}$, where
$N_{\mbox{spin}}$ is the number of spins in the primary cell.

We determined the properties of the two-dimensional compressible Ising
lattice defined by Eq.(\ref{eq:iham}) using Monte Carlo simulations
in the isothermal-isobaric ensemble.  The
simulations consisted of two-million moves without data accumulation
followed by 10 million Monte Carlo moves with the accumulation of data
at each temperature.  Changes in the $R$-parameter were made forty per
cent of the time and the two-dimensional pressure was arbitrarily set to
1 atomic unit of pressure.  Simulations were performed on 4x4, 8x8,
16x16 and 32x32 lattices with periodic boundary conditions included.

Shown in Figure \ref{f:cpi} is the constant pressure heat capacity of the
compressible Ising
model for  4x4, 8x8, 16x16 and 32x32 lattices from top to bottom in
the figure.  The peak at 240K in the heat capacity occurs at the same
temperature as rapid changes in the magnetization of the model.
As the number of spins in the simulation is increased the width
of the peak narrows as expected.  Given in Figure \ref{f:ki} is the isothermal
compressibility as a function of temperature.  Again from top to bottom
is $\kappa$ for a 4x4, 8x8, 16x16 and 32x32 lattice.  For the
4x4 lattice, no peak appears in the compressibility.  As the sample size
is increased, a peak develops in the compressibility at the transition
temperature.  Evidently,
the existence of a peak in
the compressibility is more sensitive to finite
size effects than the heat capacity.  Although not shown here, the
isobaric coefficient of thermal expansion shows a peak at the transition
temperature
for all lattice
sizes.  These Ising results lend support to our assumption that the lack
of peak in $\kappa$ in the isothermal-isobaric simulations of ammonium
chloride is a result of finite size effects.

\section{Summary and discussion}

In this work we have applied Monte Carlo methods in the canonical and
isothermal-isobaric ensembles to calculate the thermodynamic properties
of $\nhfcls$ modeled by simple point charge pair interaction potentials.
In both ensembles the model potential predicts an order to disorder
transition associated with the rotational orientations of the ammonium
ions in the lattice.  In the isothermal-isobaric ensemble at 1
atmosphere pressure, the transition temperature is about 250K,
as determined from the rapid growth of the susceptibility.
  Although
we have made no effort to determine the transition temperature with any
precision, the agreement between the approximately determined
temperature and the experimental result (243K) is satisfying.

The central sample cell used in the simulations contained a maximum of
27 ion pairs.  Extensions to larger central cell sizes were inhibited by
the large portion of the computer time needed to evaluate the Ewald
corrections.  Apparently this size limitation produced unreliable values
for the isothermal compressibility, a quantity that compressible Ising
calculations imply is sensitive to the size of the central cell.

There are several outstanding questions about the behavior of ammonium
chloride that require further attention.  We found no evidence in our
simulations for a rapid change in the molar volume at the transition
temperature.  Since the transition is known to be first order at
atmospheric pressure\cite{WeinGarl},
it is worth examining some of the physical effects
not included in the simulations.  A true discontinuity in the lattice
constant would require an infinite central simulation cell.  The unseen
rapid change in volume may be another example of a finite size effect.
However, in the simulations presented here we have included only those
low frequency vibrational modes where the overall crystal symmetry is
unaltered.  Calculations that include more general variations in the
locations of the ionic mass centers would clearly be of interest.
Inclusion of the internal vibrations on the ammonium ions can be
expected to be of less importance owing to the high frequencies of those
motions.

Another approximation in the results has been the application of classical
mechanics.  Classical calculations can be interpreted as the infinite
mass limit of a corresponding quantum calculation.
If quantum effects were not important, our simulations would be equally
applicable to the phase diagram of $\ndfcls$.
Experimentally,
a
significant shift of the tricritical point to lower pressures is
known to occur when the
ammonium ions are deuterated~\cite{NisPoy,WeinGarl}.
It is possible that the classical system at
atmospheric pressure has a
continuous disordering transition,
 and the location (or existence) of a tricritical point is a consequence of
quantum effects.  A quantum path integral study of ammonium chloride
within the same model potential would clearly be of interest.

\section*{Acknowledgements}

We would like to thank Professors P. Nightingale, R. Stratt and J. Doll
for helpful discussions.  RQT would like to thank Dr. Michael New
and the members of the Ohio Supercomputer Center
 Computational Chemistry Electronic Forum (chemistry@osc.edu)
 for
helpful e-mail conversations regarding simulation methods.
Acknowledgement is made to the Donors of the Petroleum Research Fund of
the American Chemical Society for support of this work.  The
computational work reported here was supported by an equipment grant
from the National Science Foundation(CHE-9203498).

\newpage
\begin{table} \centering
\caption{The parameters used in the model potential$^d$}

\vspace{.15in}
\renewcommand{\arraystretch}{1.5}
\begin{tabular}{*{5}{c}}
Pair & $A_{ij}$ & $\alpha _{ij}$ &  $C_6$ & $D_{12}$  \\
\hline
H-H$^a$ & 1.0162 &  1.9950 & 2.9973& 0 \\
N-N$^b$ & 104.74 & 1.5611 & 25.393 & 0 \\
Cl-Cl$^b$ & 125.55 & 1.7489 & 113.68 & 0 \\
H-N$^a$ & 10.318 & 1.7780 & 8.7229 & 0 \\
H-Cl$^c$ & 0 & 0 & 10.033 & 43884.0 \\
N-Cl$^a$ & 114.22 & 1.6550 & 53.736 & 0 \\
\end{tabular}
\end{table}
\parbox[b]{10in}{$^a$ Combining rules \\
$^b$ Reference \onlinecite{KleinMacOz} \\ $^c$ Reference \onlinecite{pr}
\\$^d$ Units of energy in Hartree and units of length in Bohr}

\newpage
\section*{Figure Captions}
\begin{enumerate}
\item
The two ``spin'' orientations of an ammonium ion in a cage of 8 chloride ions.
  Each
orientation represents a potential minimum with the hydrogen atoms on
the ammonium ion pointing to the shaded chloride ions.  The two
orientations are related
to one another by rotating the ammonium ion by an angle of
$\pi /2$ about the $z$-axis.  The arrows
between the hydrogen atoms and the chloride ions are included for
clarity.\cite{xmol}
\item
The potential energy $U$ in kJ mol$^{-1}$of an
oriented ammonium chloride lattice with 27 ion
pairs in the
central cell and Ewald corrections included.  The energy is given as a
function of the rotation angle $\phi$ for a central ammonium ion as it
is rotated out of orientation with the remaining ammonium ions in the
lattice (see Figure 1).
\item
The difference in energy $\Delta U$
in kJ mol$^{-1}$ between an ammonium chloride lattice with
with all ammonium ions oriented in the same direction except the
central ion and an ammonium chloride lattice with all ammonium ions
oriented in the same direction except for the central ion and another
ion a distance $R$ in \AA \ from the central ion.  The rapid approach to an
asymptote is a consequence of the octapole-octapole interaction and
lends support to the representation of the lattice by a nearest
neighbor Ising model.
\item
The average potential energy
in kJ mol$^{-1}$of NH$_4$Cl using 8 ion pairs in the
central simulation cell as a function of temperature.
Both curves were generated from an initial
configuration with the ammonium ions in a random rotational configuration.
In the
upper curve ordinary Metropolis moves were used whereas in the lower
curve a magnified step size was included ten percent of the time.  The
quasiergodicity problems evident in the upper curve at the
lower temperatures are removed by the
``magwalking'' strategy.  The error bars are smaller than the plotted
points.
\item
An 8 ion pair representation of the ammonium chloride lattice.  The
representation is a particular configuration taken from a canonical
simulation at 300K.  The arrows attached to each ammonium ion represent
the algebraic sign of $M_j$.  The up arrows are on ammonium ions with
positive $M_j$ and the down arrows are on ammonium ions with negative
$M_j$.  The arrows represent a mapping of the ammonium ion orientation
onto a spin variable.\cite{xmol}
\item
The constant volume heat capacity per ion pair
in units of $k_B$
 of ammonium chloride as a function of
temperature.  The data in the upper curve were obtained from canonical
simulations
with 8 ion pairs in the central cell, and the data in the lower curve were
obtained from canonical simulations with 27 ion pairs in the central cell.
By increasing the size of the central simulation cell the heat capacity
narrows with a peak maximum at 210K.
The error bars are at the double standard deviation level.\label{f:cv}
\item
The square of the magnetization as a function of temperature for a 27
ion pair representation of the ammonium chloride lattice in the canonical
ensemble.  The points
represented by diamonds were computed directly using Eq.(\ref{eq:mj}).
The points represented by triangles were obtained by mapping each ammonium
ion orientation onto a spin variable as discussed in the text.  The
magnetization changes rapidly at the same temperature that the heat
capacity shows a maximum in Figure \ref{f:cv}.  The error bars in this
figure are smaller than the representations of the plotted points.
\label{f:m2}
\item
The susceptibility as a function of temperature for a 27 ion pair
representation of the ammonium chloride lattice in the canonical
ensemble.  The error bars are at the double standard deviation level.
The maximum in the peak of the susceptibility matches the maximum in the
heat capacity and the region of rapid change in the magnetization.
\label{f:chi}
\item
The square of the magnetization as a function of temperature for a 27
ion pair representation of the ammonium chloride lattice in the
isothermal-isobaric
ensemble at a pressure of 1 atmosphere.  The points
represented by diamonds were computed directly using Eq.(\ref{eq:mj}).
The points represented by triangles were obtained by mapping each ammonium
ion orientation onto a spin variable as discussed in the text.  The
magnetization changes rapidly at the same temperature that the heat
capacity and isobaric coefficient of thermal expansion
have maxima.  The error bars in this
figure are smaller than the representations of the plotted points.
\label{f:m2npt}
\item
The susceptibility as a function of temperature for a 27 ion pair
representation of the ammonium chloride lattice in the
isothermal-isobaric
ensemble.  The error bars are at the double standard deviation level.
The maximum in the peak of the susceptibility  identifies the onset of
the order to disorder transition.
\label{f:chinpt}
\item
The lattice parameter $a_0$ in \AA \ as a function of temperature for the
27 ion
pair representation of the ammonium chloride crystal at 1 atmosphere
pressure calculated in the isothermal-isobaric ensemble.  Although no
discontinuous change in the lattice parameter is seen, a slope change is
apparent at the transition temperature.  The inflection matches the
location of the peak in the isobaric coefficient of thermal expansion.
The error bars are at the double standard deviation level.\label{f:a0npt}
\item
The constant pressure heat capacity $C_p$ per ion pair
in units of $k_B$, the isobaric
coefficient of
thermal expansion $\alpha$ in K$^{-1}$and the
isothermal compressibility $\kappa$ in atmospheres$^{-1}$
as a function of temperature for the 27 ion pair representation of the
ammonium chloride crystal at 1 atmosphere pressure in the
isothermal-isobaric ensemble.  The maxima in $C_p$ and $\alpha$ match
the temperature at which there are rapid changes in the order parameter.
The lack of a maximum in $\kappa$ is likely a result of finite size
effects (see text).
The error bars are at the double standard deviation level.\label{f:cpnpt}
\item
The constant pressure heat capacity for the compressible Ising model.
The four panels from top to bottom are results for 4x4, 8x8 16x16 and
32x32 lattices with periodic boundary conditions.
The error bars are at the double standard deviation level.\label{f:cpi}
\item
The isothermal compressibility in inverse atomic units of pressure
 for the compressible Ising model.
The four panels from top to bottom are results for 4x4, 8x8 16x16 and
32x32 lattices with periodic boundary conditions.
The error bars are at the double standard deviation level.\label{f:ki}
\end{enumerate}

\begin{references}
\bibitem{expspin}
When the hydrogen atoms point directly to the chloride ions in any cubic
cage, there are 24 possible orientations.  The 24 orientations consist
of two sets each set of which has 12 equivalent configurations.  In the
Ising picture the order to disorder transition corresponds to the
spontaneous breaking of the symmetry that takes one set of 12
configurations to the other set of 12 configurations.
\end{references}
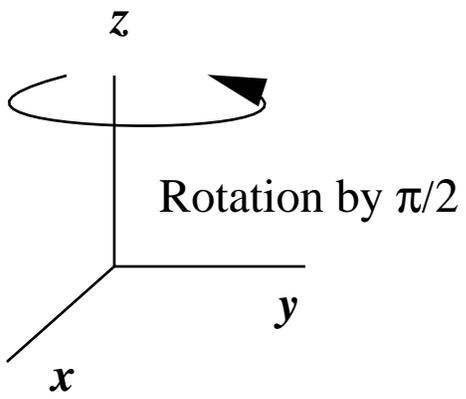
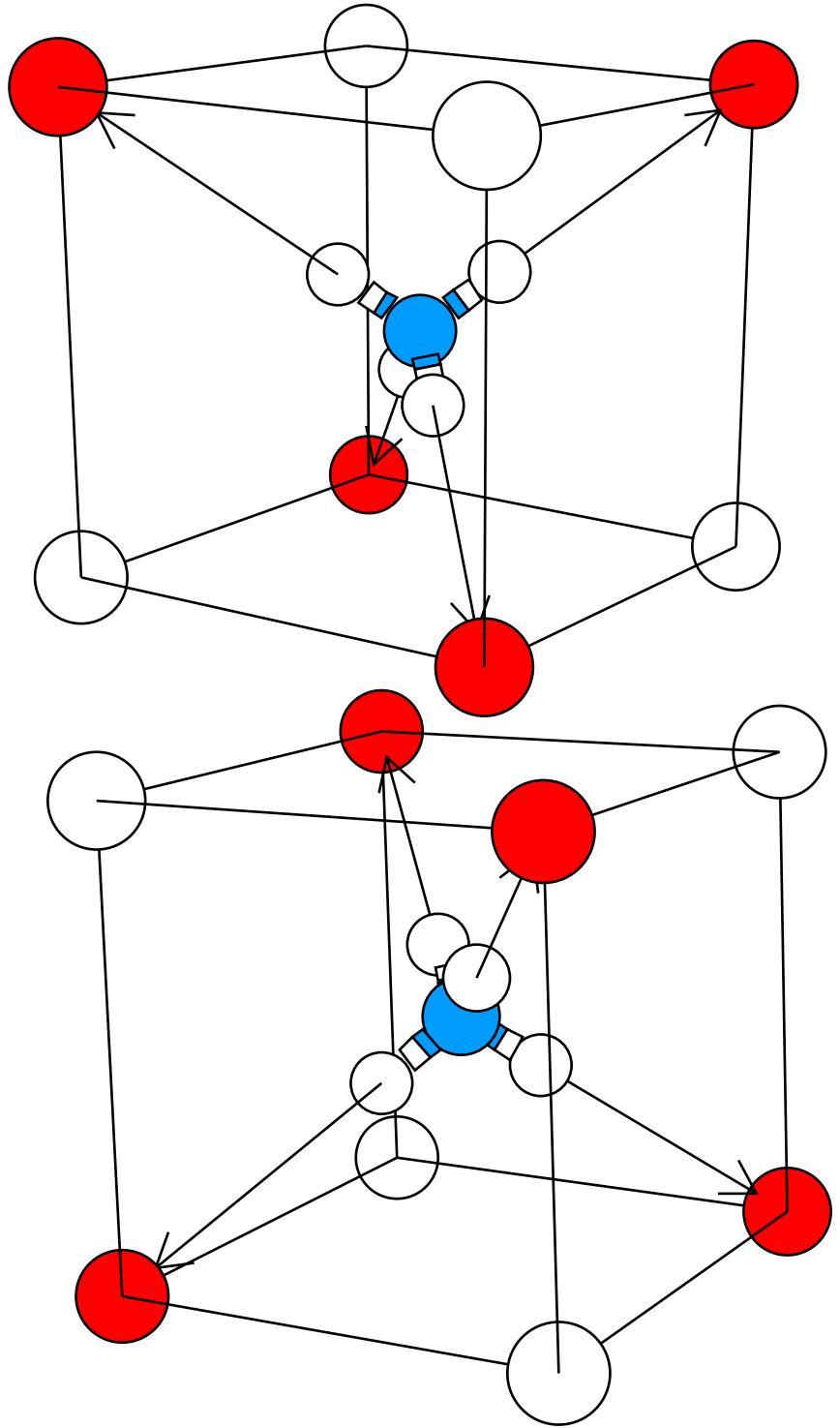
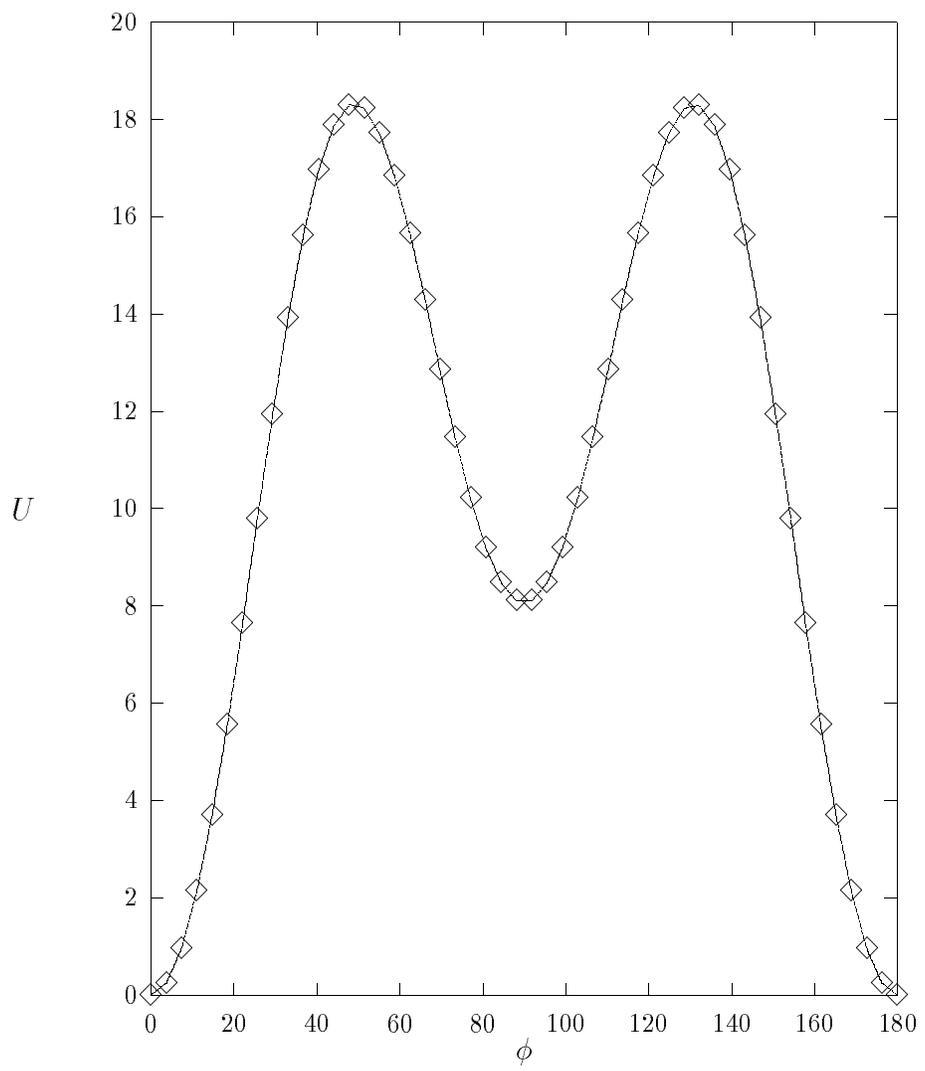
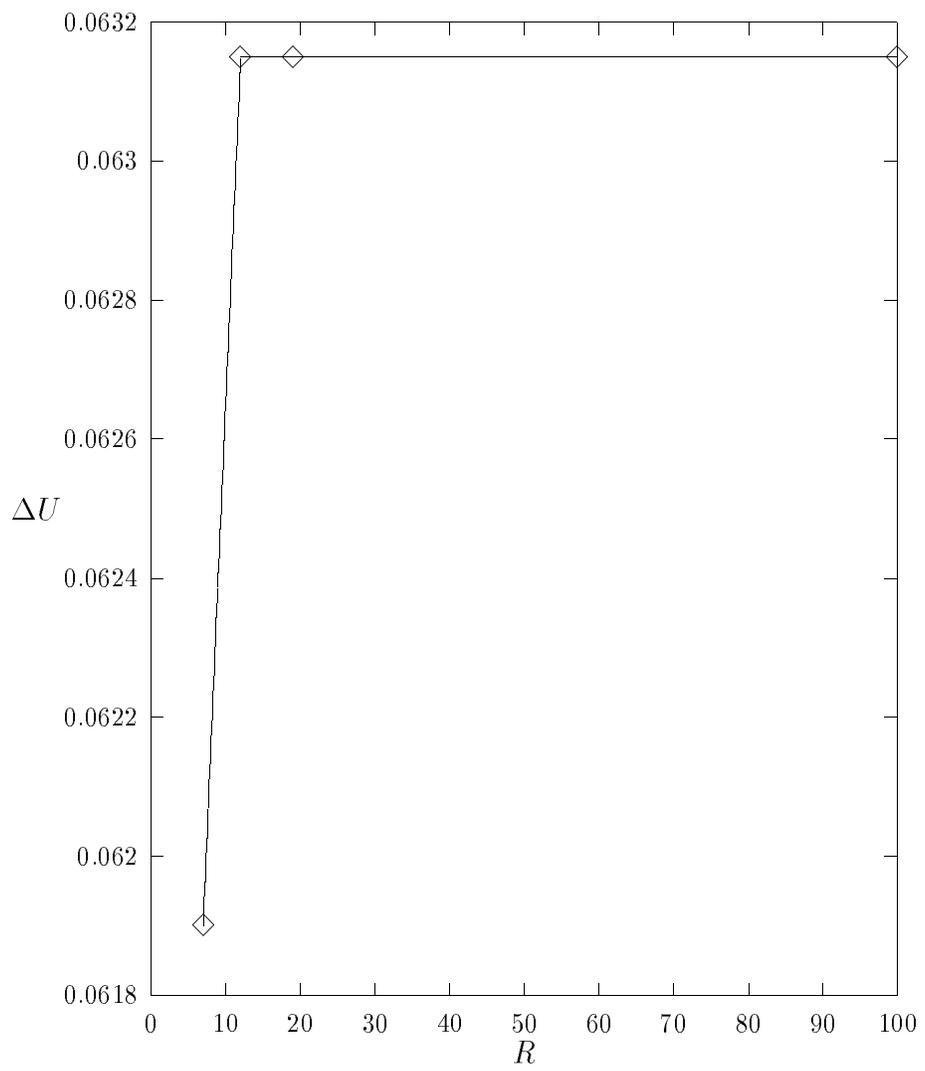
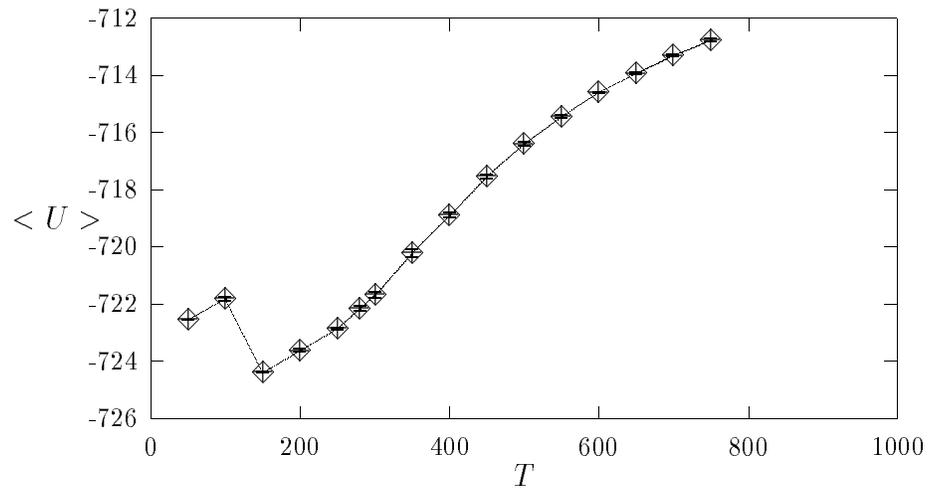
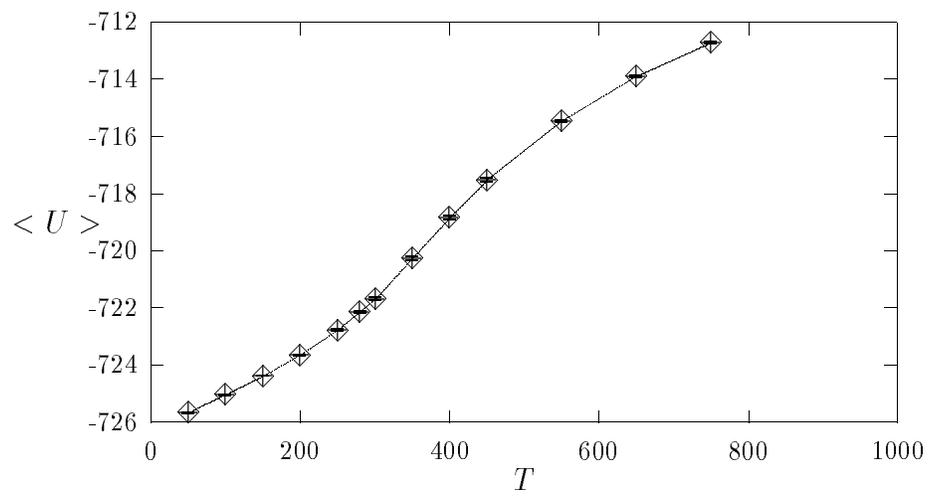
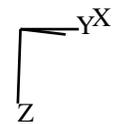
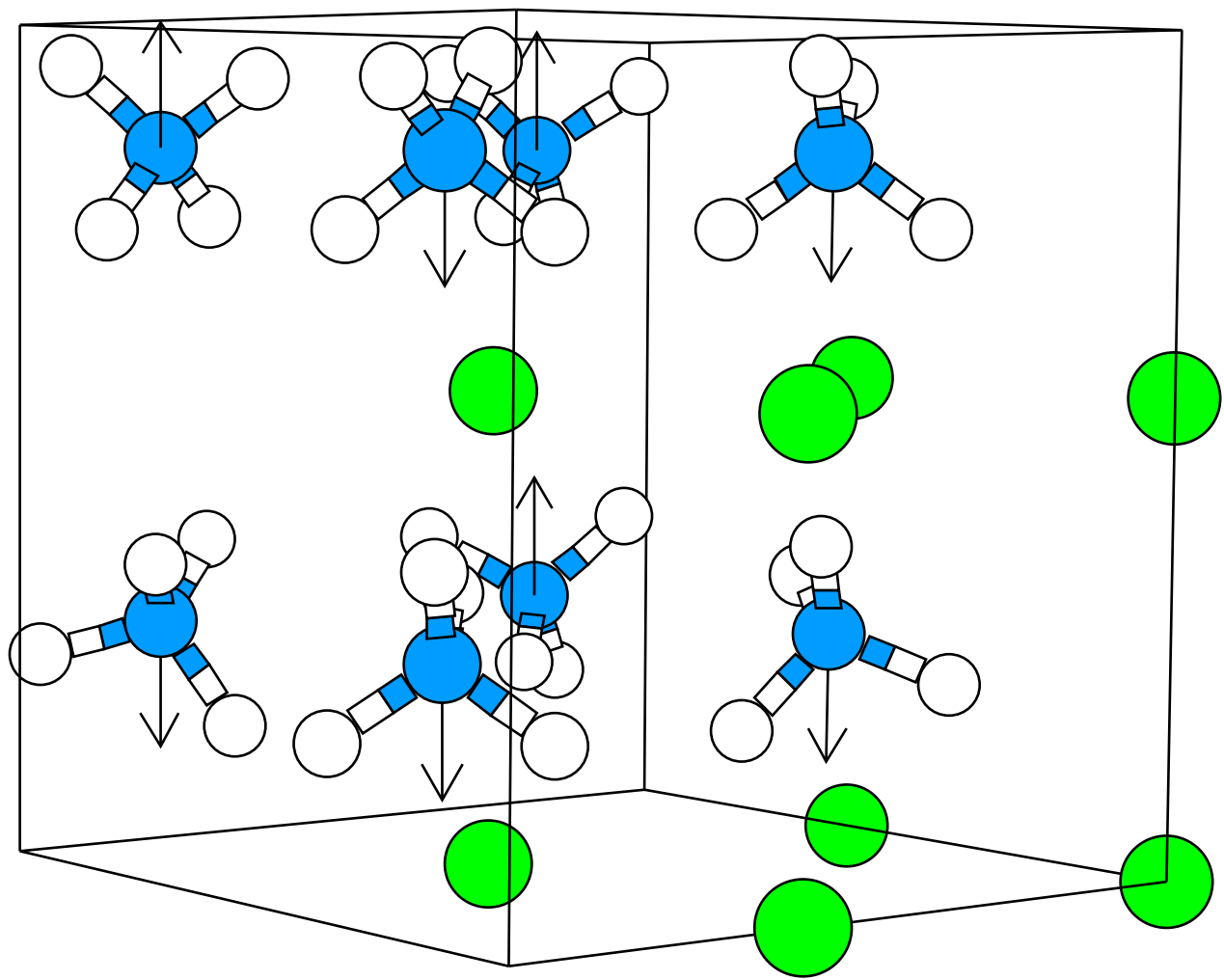
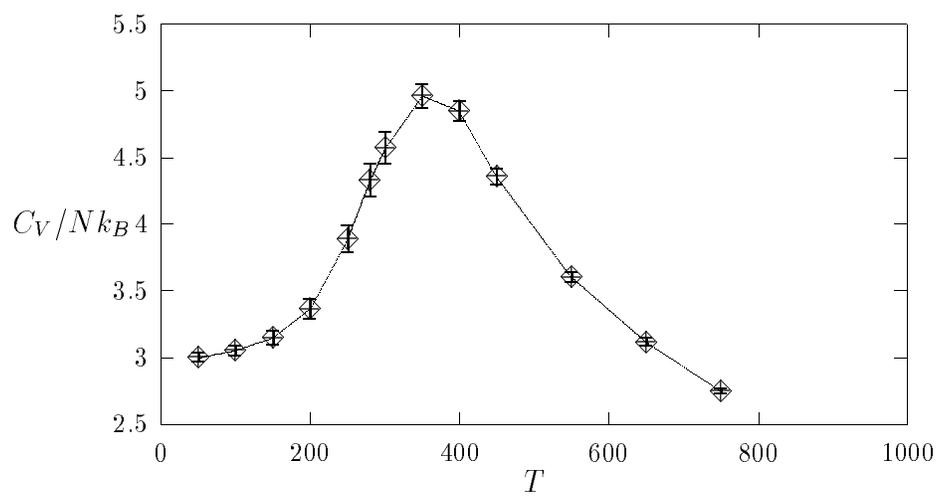
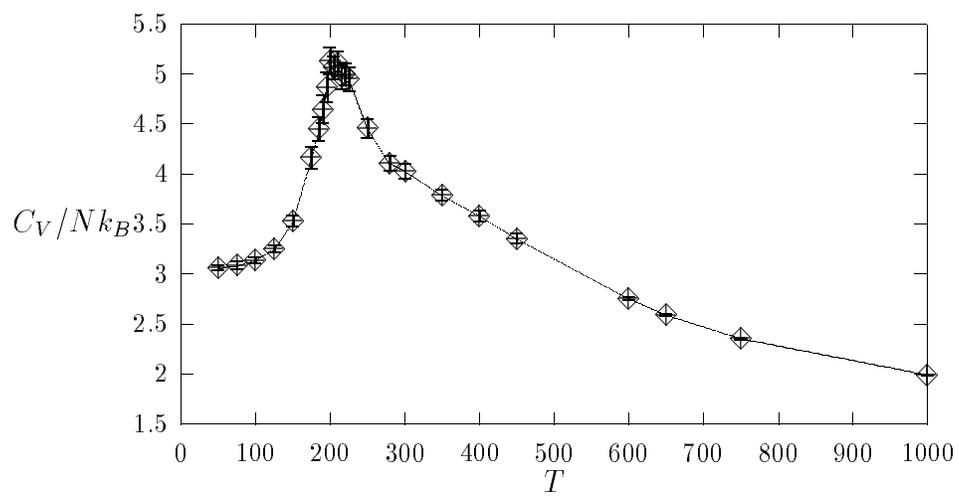
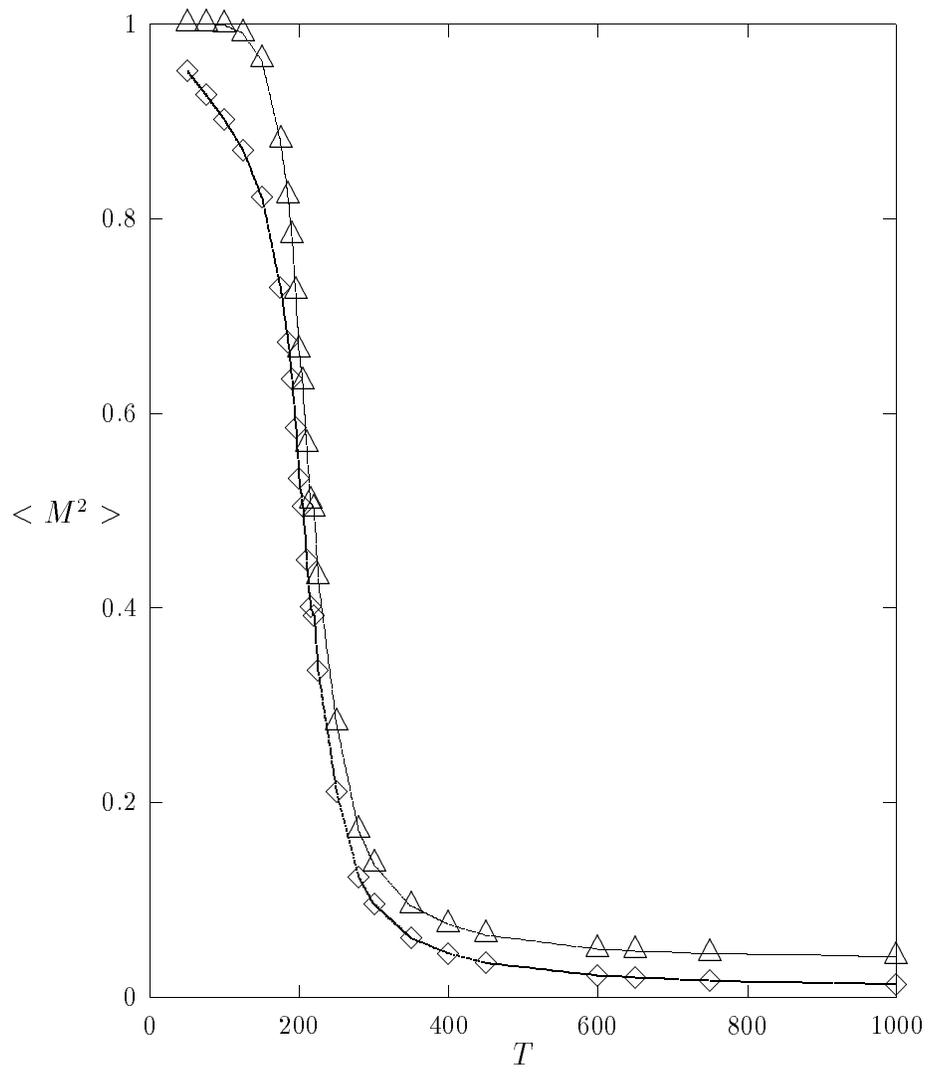
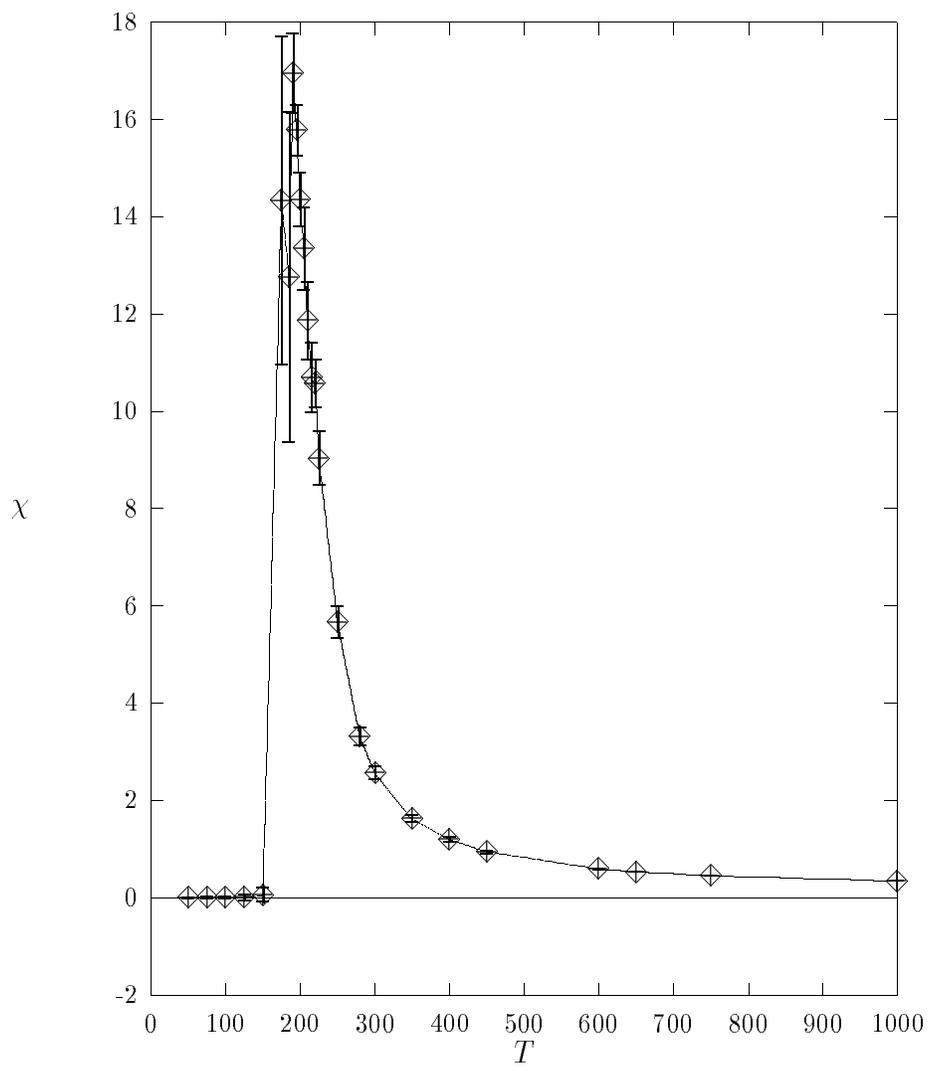
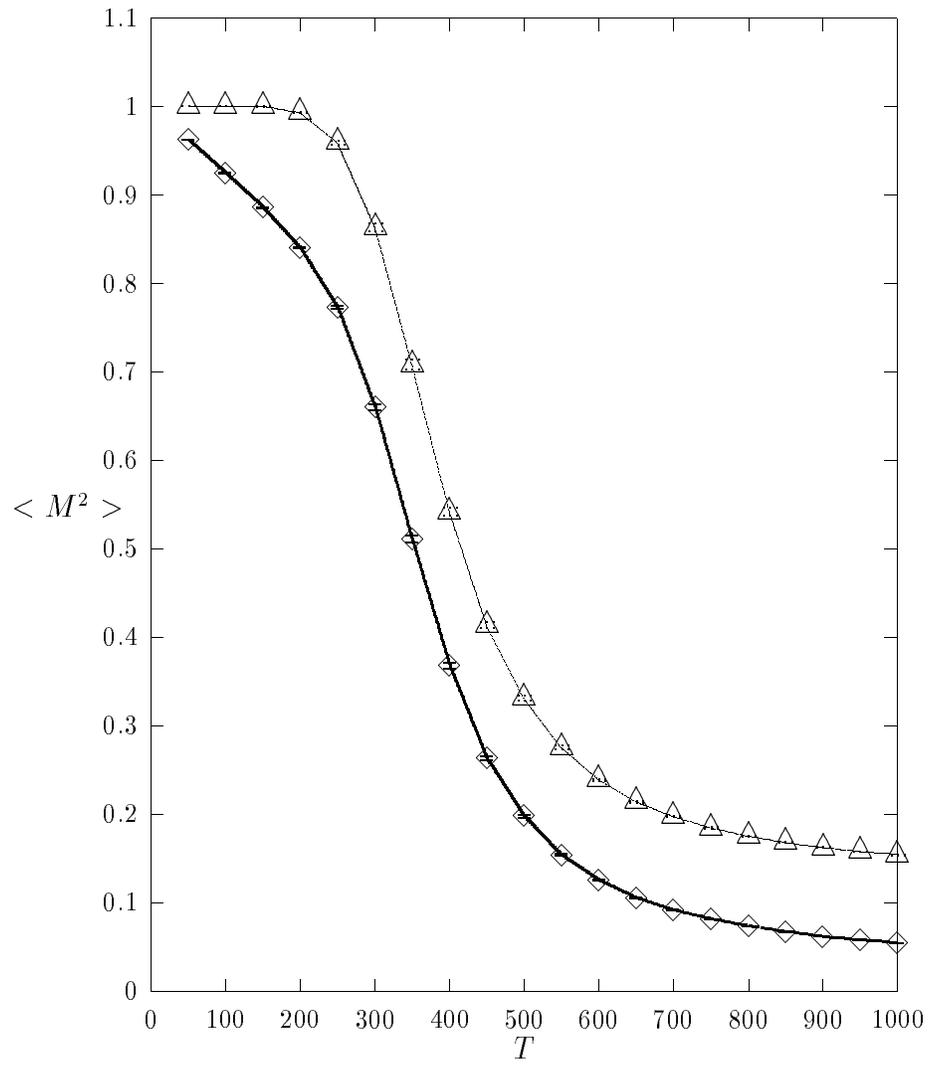
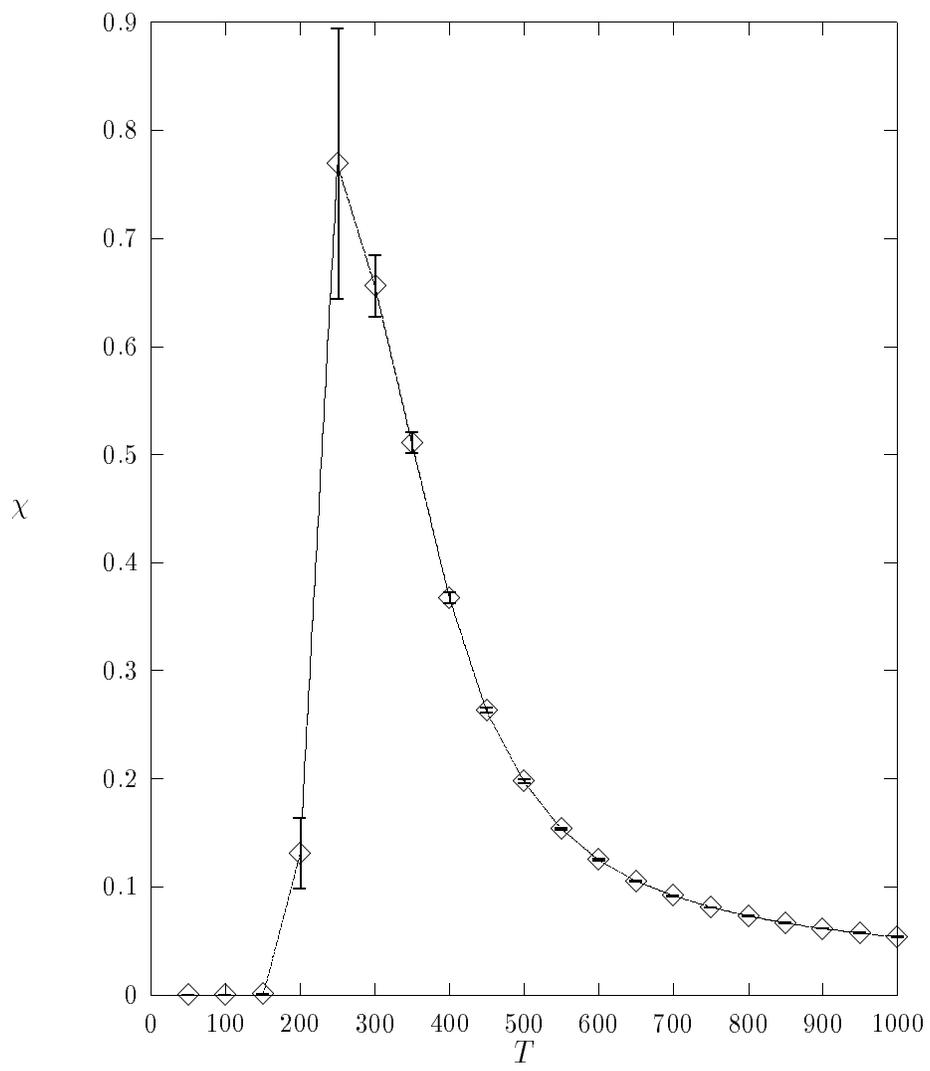
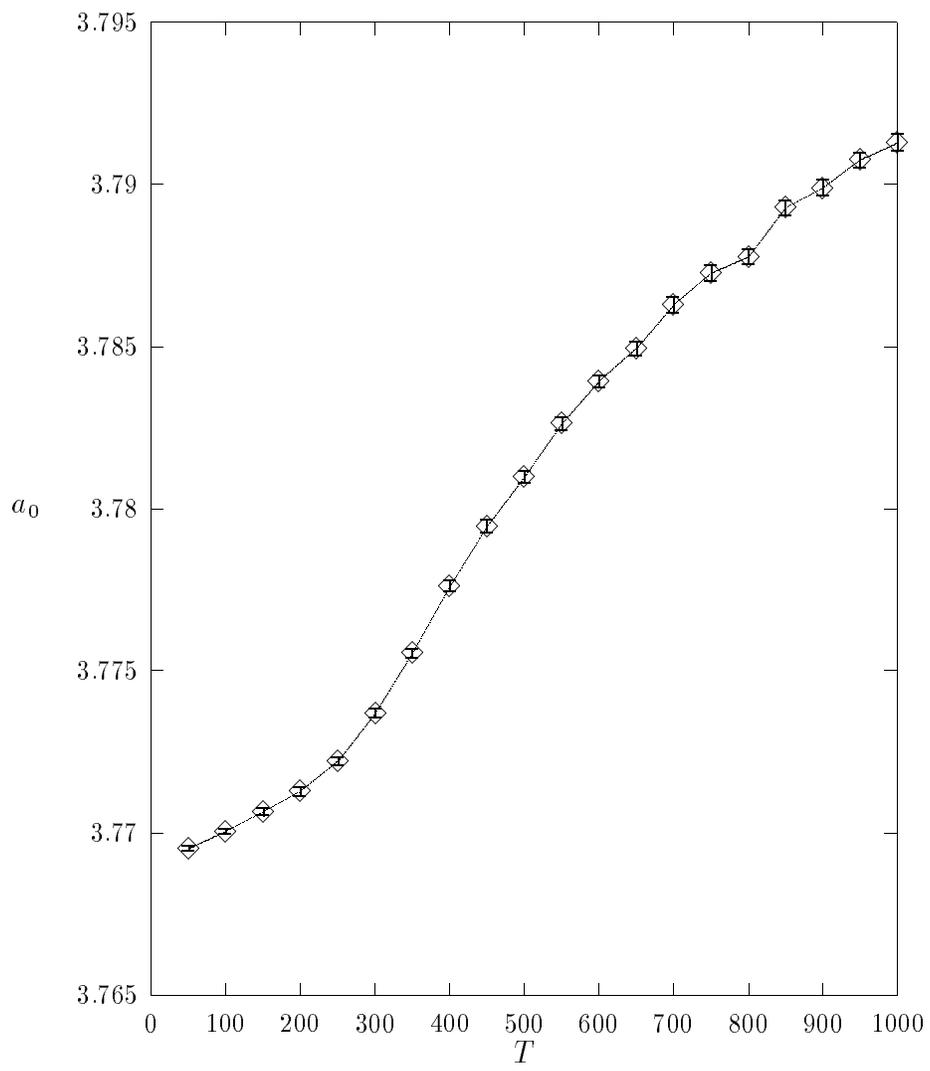
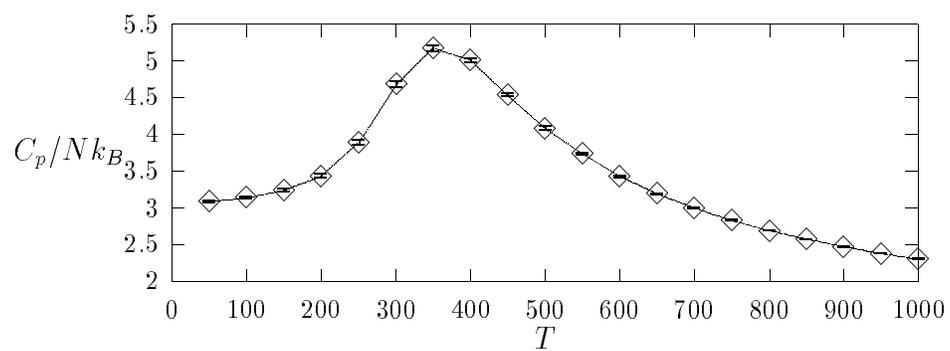
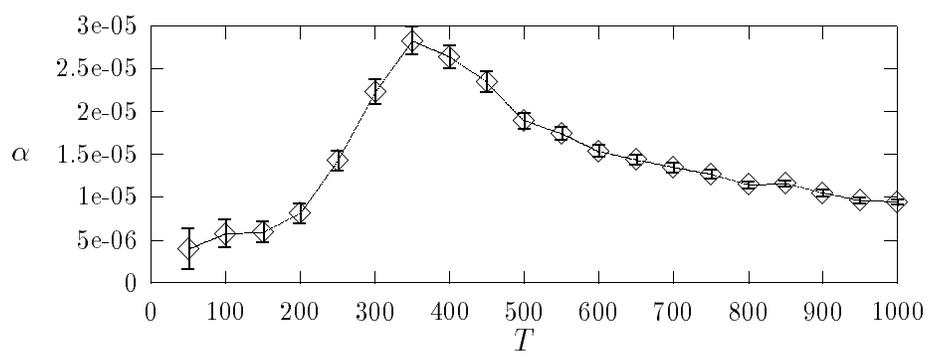
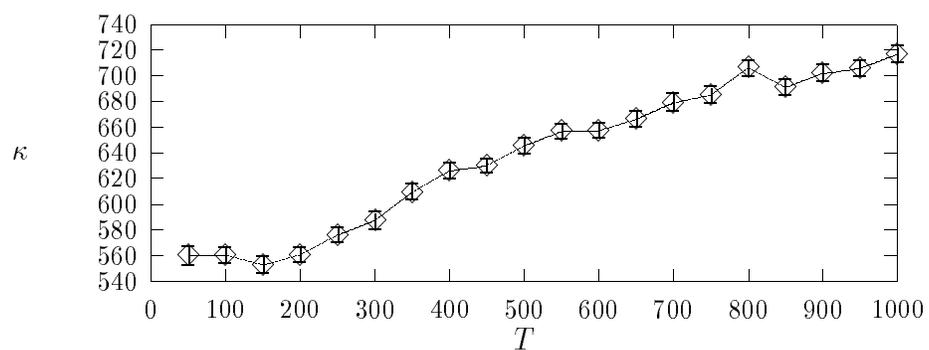
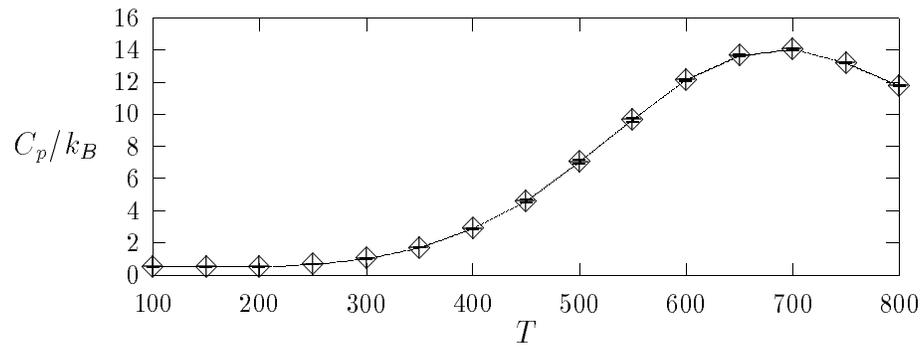
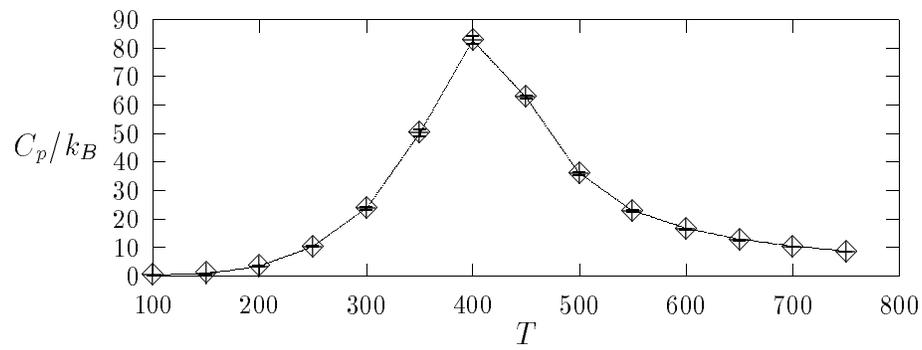
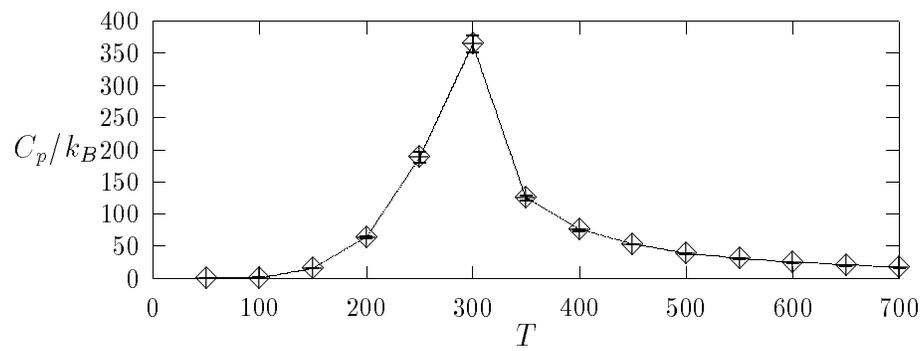
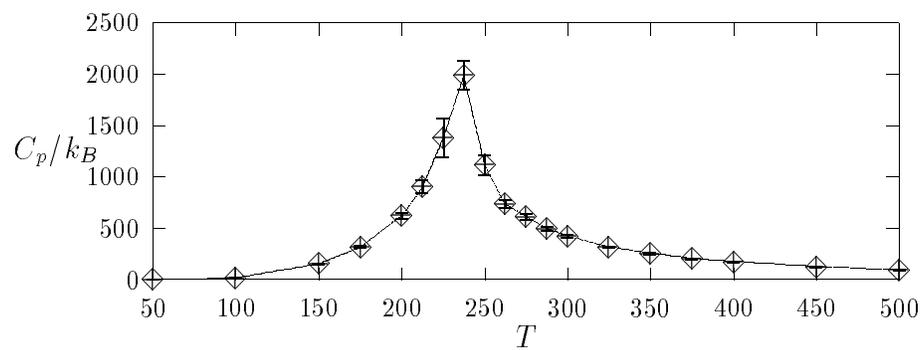
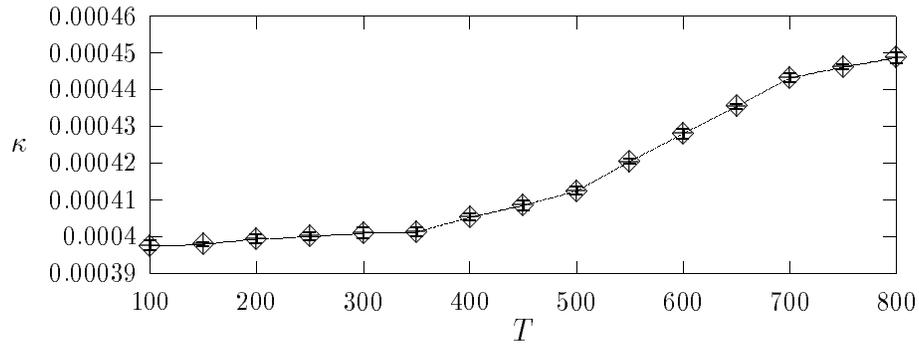
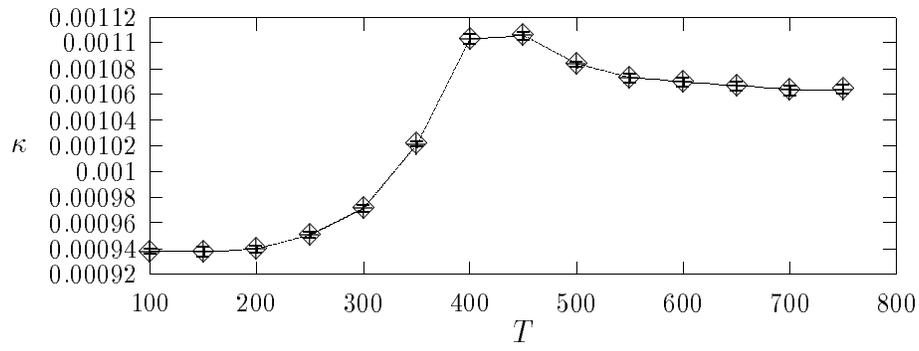
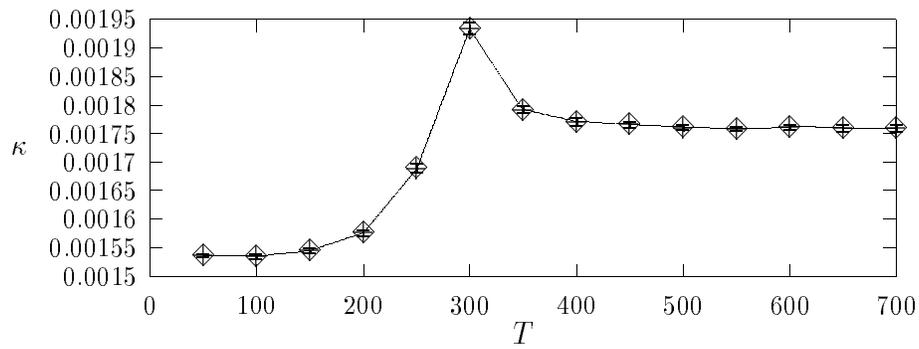
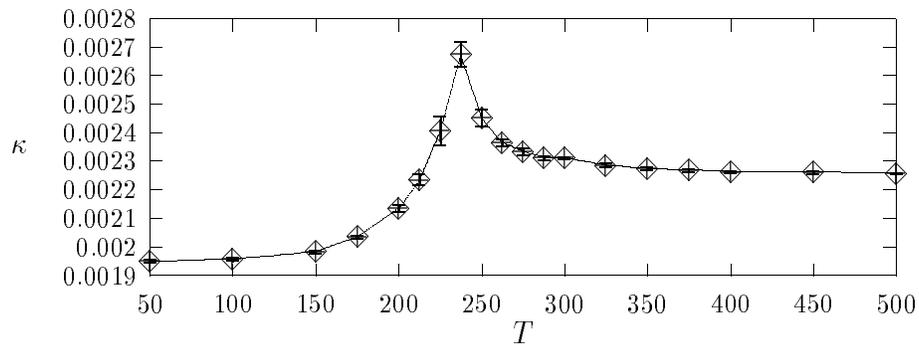
\end{document}